\documentclass[aps,nofootinbib,longbibliography,twocolumn]{revtex4-1}

\usepackage{amsmath}
\usepackage{amssymb} 
\usepackage[cm]{fullpage}
\usepackage{graphicx} 
\usepackage{siunitx}
\usepackage{tabularx} 
\usepackage[colorlinks,linkcolor=blue,citecolor=blue,urlcolor=blue]{hyperref}
\usepackage[nameinlink]{cleveref}
\usepackage[table,xcdraw]{xcolor}
\usepackage[utf8]{inputenc}
\usepackage{tabularx}
\usepackage{verbatim}

\DeclareSIUnit\parsec{pc}

\newcommand{\be}{\begin{equation}}
\newcommand{\ee}{\end{equation}}
\newcommand{\beqx}{\begin{equation*}}
\newcommand{\eeqx}{\end{equation*}}
\newcommand{\beqa}{\begin{eqnarray}}
\newcommand{\eeqa}{\end{eqnarray}}
\newcommand{\beqax}{\begin{eqnarray*}}
\newcommand{\eeqax}{\end{eqnarray*}}

\newcolumntype{C}[1]{>{\centering\arraybackslash}p{#1}}

\begin{document}

\title{Constraints on Late Time Violations of the Equivalence Principle in the Dark Sector}
\date{\today}
\author{Cameron C. Thomas and Carsten van de Bruck}

\affiliation{Consortium for Fundamental Physics, School of Mathematics and Statistics, University of Sheffield, Hounsfield Road, Sheffield S3 7RH, United Kingdom}

\begin{abstract}
If dark energy is dynamical due to the evolution of a scalar field, then in general it is expected that the scalar is coupled to matter. While couplings to the standard model particles are highly constrained by local experiments, bounds on couplings to dark matter (DM) are only obtained from cosmological observations and they are consequently weaker. It has recently been pointed out that the coupling itself can become non-zero only at the time of dark energy domination, due to the evolution of dark energy itself, leading to a violation of the equivalence principle (EP) in the dark sector at late times. In this paper we study a specific model and show that such late-time violations of the EP in the DM sector are not strongly constrained by the evolution of the cosmological background and by observables in the linear regime (e.g. from the cosmic microwave background radiation). A study of perturbations in non-linear regime is necessary to constrain late--time violations of the equivalence principle much more strongly. 
\end{abstract}

\maketitle

\section{Introduction}
\label{Introduction}

There are several reasons to study alternative theories to the cosmological constant as a model for dark energy (DE). Firstly, if DE is due to a non-vanishing cosmological constant, its value has to be very small to fit the data. The expected theoretical value, however, is much larger. This problem has not been solved, but there are attempts to address this problem (see \cite{Weinberg:1988cp,Padmanabhan:2002ji,Martin:2012bt} and references therein). Secondly, there are several tensions between data sets, providing  tantalising hints that the standard model of cosmology, the $\Lambda$-Cold-Dark-Matter($\Lambda$CDM) model, may be in need of corrections (we refer to \cite{Abdalla:2022yfr} for a recent overview over the tensions and \cite{DiValentino:2021izs} for an overview of attempts to solve the tensions). Among the extensions of the $\Lambda$CDM model which remain the best motivated ones are scalar-field models of DE, in which the accelerated expansion is driven by a scalar field \cite{Wetterich:1987fm,Ratra:1987rm,Caldwell:1997ii}. It is expected that, in general, the scalar is coupled to at least one species of matter, unless there is a symmetry which forbids such couplings \cite{Carroll:1998zi}. Such a coupling results in an additional force mediated between the coupled species. Since the interaction between DE and ordinary matter is strongly constrained, in some models only the coupling to cold dark matter (CDM) is of cosmological significance (see e.g.\cite{Wetterich:1994bg,Amendola:1999er,Farrar:2003uw,VanDeBruck:2017mua,Archidiacono:2022iuu} and references therein). It is this type of theory which we consider in this paper. 

It has recently been suggested, that the potential energy of scalar fields appearing in string theory cannot be arbitrarily flat \cite{Obied:2018sgi,Agrawal:2018own}, see \cite{Palti:2019pca} for an overview of the swampland programme. If true, the accelerated expansion cannot be driven by a cosmological constant (de Sitter space is not realised in string theory) and the equation of state of dark energy is not constant and deviates potentially significantly from the value expected in the $\Lambda$CDM model. Additionally, a coupling of the scalar to some sectors in the theory are expected. Based on these observations, several phenomenological models have been proposed recently \cite{van_de_Bruck_2019,Agrawal:2019dlm,Olguin-Trejo:2018zun,Brax:2019rwf,ValeixoBento:2020ujr,Brinkmann:2022oxy}. In this paper we study specific a model in which the coupling function between the dark energy field and dark matter has a mini\-mum \cite{van_de_Bruck_2019}. As a result of the mini\-mum, the coupling switches on only at late times, at the beginning of the dark energy dominated epoch. One of our main results of this paper is that the regime in which linear perturbation theory is valid does not constrain the parameter of the model greatly. In other words, late time violations of the equivalence principle in the dark sector are not strongly constrained by studying the background evolution or CMB anisotropies. Instead, to obtain stronger constraints a study of the non-linear regime in considerable detail is needed. 

The paper is organised as follows. In \Cref{InteractingDarkEnergy} we present the model and its parameter. In \Cref{MethodologyDataandResults} we describe our methodology, describe the data sets used and present the constraints on the model. We conclude in \Cref{Conclusions}.

\section{Interacting Dark Energy} 
\label{InteractingDarkEnergy}

The model we consider consists of the gravitational sector described by the Einstein--Hilbert action without cosmological constant, a part which describes the standard model (SM) particles and a part for DE described by a scalar field $\phi$ with potential energy $V(\phi)$. Finally, the interaction between DE and DM is described by a conformal coupling. The full action reads 

\begin{eqnarray}
{\cal S} &=& \int d^4 x \sqrt{-g}\left( \frac{M_{\rm Pl}^2}{2} {\cal R} - \frac{1}{2}g^{\mu\nu} \partial_\mu \phi \partial_\nu\phi - V(\phi) \right)\nonumber \\ 
&+& \int d^4 x \sqrt{-g} {\cal L}_{\rm SM}(g,\Psi_i)
+\int d^4 x \sqrt{-{\tilde{g}}} {\cal L}_{\rm DM}({\tilde g},\chi) \label{eq:action}
\end{eqnarray}
where $M_{\rm Pl}$ is the reduced Planck mass, ${\cal R}$ is the Ricci--scalar, the SM fields are denoted by $\Psi_i$ and $\chi$ is the DM field (assuming here for simplicity that dark matter consists of only one species). The metrics $g$ and $\tilde g$ are related by a conformal transformation $\tilde{g}_{\mu\nu} = C(\phi) g_{\mu\nu}$. Such theories have been discussed in considerable length in the literature, but the new ingredient in this paper is that the function $C(\phi)$ has a minimum. That the coupling functions in string theory could possess a minimum due to non-perturbative effects was suggested in the works by Damour and Polyakov \cite{Damour:1994zq} and has been used in \cite{Brax:2010gi} to construct a dilaton-model of DE. 
Combining these theoretical developments motivated us in \cite{van_de_Bruck_2019} to consider a specific model in which $C(\phi)$ has a minimum at some value scalar field value $\phi_*$. To be concrete, in this paper we consider the following function for $C(\phi)$:
\be
C(\phi) = \cosh{\left( \sqrt{\alpha} \, (\phi - \phi_*)/M_{\rm Pl}  \right), }
\ee
where $\phi_*$ denotes the minimum of the function $C(\phi)$ and $\alpha$ is a constant. In this paper we choose $\phi_* = 1~M_{\rm Pl}$ without loss of generality. In \cite{van_de_Bruck_2019} it was pointed out that even if the field starts away from the minimum in the very early universe, there are attractor mechanisms at work in the early universe which drive the field towards the minimum quickly. Nevertheless, in our analysis below we allow the field to start away from the minimum value at $\phi_*$ to find constraints on the initial conditions of the DE field. In our analysis we choose an exponential potential with 
\be
V(\phi) = V_0 e^{-\lambda \phi/ M_{\rm Pl}}, 
\ee
where $\lambda$ denotes the slope of the potential, which in string theory according to \cite{Obied:2018sgi} cannot be arbitrarily small and should be ${\cal O}(1)$. Finally, we assume in the following that the universe is spatially flat.

\begin{figure}[h]
    \centering
    \includegraphics[width=0.5\textwidth]{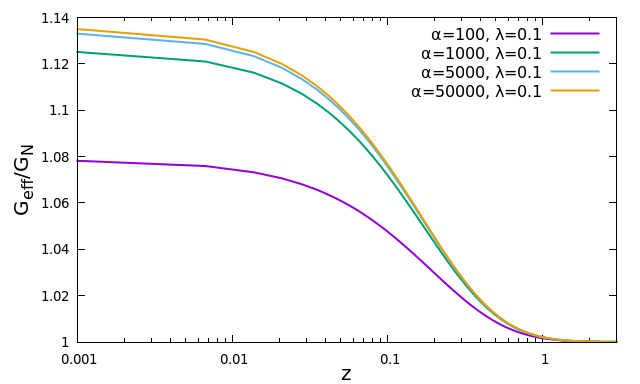}
    \includegraphics[width=0.5\textwidth]{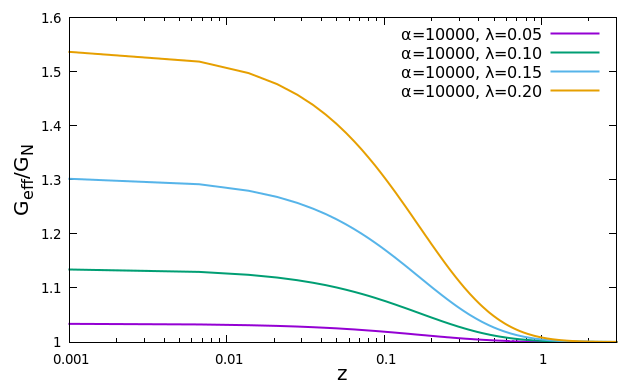}
    \caption{The evolution of
the effective gravitational constant, defined in \Cref{eqn:Geff}, for models with a different value of $\alpha$ but same value of $\lambda = 0.1$ (top) and for models with a different value of $\lambda$ but same value of $10^{-4} \alpha = 1$ (bottom). Where applicable, the values of our cosmological parameters are taken from the best fit values of a $\Lambda$CDM cosmology based on $Planck$ TTTEEE+lowE, such as in Table 2 of \cite{Planck:2018vyg}.}
    \label{fig:Geff_plot_alpha}
\end{figure}

Because of the coupling, there is an exchange of energy between DM and the DE field. As a result, the evolution of the DM density and the modified Klein--Gordon equation are given by $$\dot \rho_{\rm DM} + 3H \rho_{\rm DM} = \beta M_{\rm Pl}^{-1} \dot\phi \rho_{\rm DM}$$ and $$\ddot \phi + 3H\dot\phi + V_\phi = -\beta M_{\rm Pl}^{-1} \rho_{\rm DM}.$$ In these equations we have defined $$\beta = \frac{M_{\rm Pl}}{2} \frac{{\rm d} \ln C}{{\rm d} \phi}.$$ The effective gravitational constant between two DM particles is given by \cite{Amendola:2003wa,van_de_Bruck_2019}
\begin{equation}
\label{eqn:Geff}
    G_{\rm eff} = G_N \left( 1 + 2\beta^2 \right).
\end{equation}
The evolution of $G_{\rm eff}$ is shown in \Cref{fig:Geff_plot_alpha} for various choices of parameters $\alpha$ and $\lambda$. In general, the additional force between DM particles due to the scalar field only becomes significant at redshifts $z<1$, when the DE field starts to evolve due to the influence of the potential, thereby displacing it from the minimum of the coupling function. We emphasize that the effective gravitational coupling (\Cref{eqn:Geff}) between DM particles depends on $\alpha$ as well as on the scalar field. 

To summarize, the parameters of the model we seek to constrain are the slope of the potential $\lambda$, the parameter $\alpha$, which is related to strength of the coupling between DM and DE, and the initial field value $\phi_{\rm ini}$ deep inside the radiation dominated epoch.

\section{Methodology, Data and Results}
\label{MethodologyDataandResults}

\begin{table}[ht!]
\caption{Flat priors for the cosmological parameters sampled in our analysis.}
\begin{tabular}{C{3.5cm} C{3.5cm}}
\hline\hline \\[-1.5ex]
Parameter & Prior \\[1ex] 
\hline \\[-1.5ex]
$\Omega_b h^2$ & $[0.005, 0.1]$\\
$\Omega_{cdm} h^2$ & $[0.001, 0.99]$\\
$100 \theta_{s}$ & $[0.5, 10.0]$\\ 
$\ln \left( 10^{10} A_s \right)$ & $[2.7, 4.0]$\\
$n_s$ &  $[0.9, 1.1]$\\
$\tau_{reio}$ & $[0.01, 0.8]$\\ 
$\lambda$ & $[0, 2]$ \\
$10^{-4} \alpha$ & $[0, 50]$\\
$\phi_{\rm ini}/M_{\rm Pl}$ & $[0, 2]$\\
\hline \\[-2ex]
\hline
\end{tabular}
\label{table:IDE_priors}
\end{table}

\begin{figure*}[t!]
    \centering
    \includegraphics[width=0.75\textwidth]{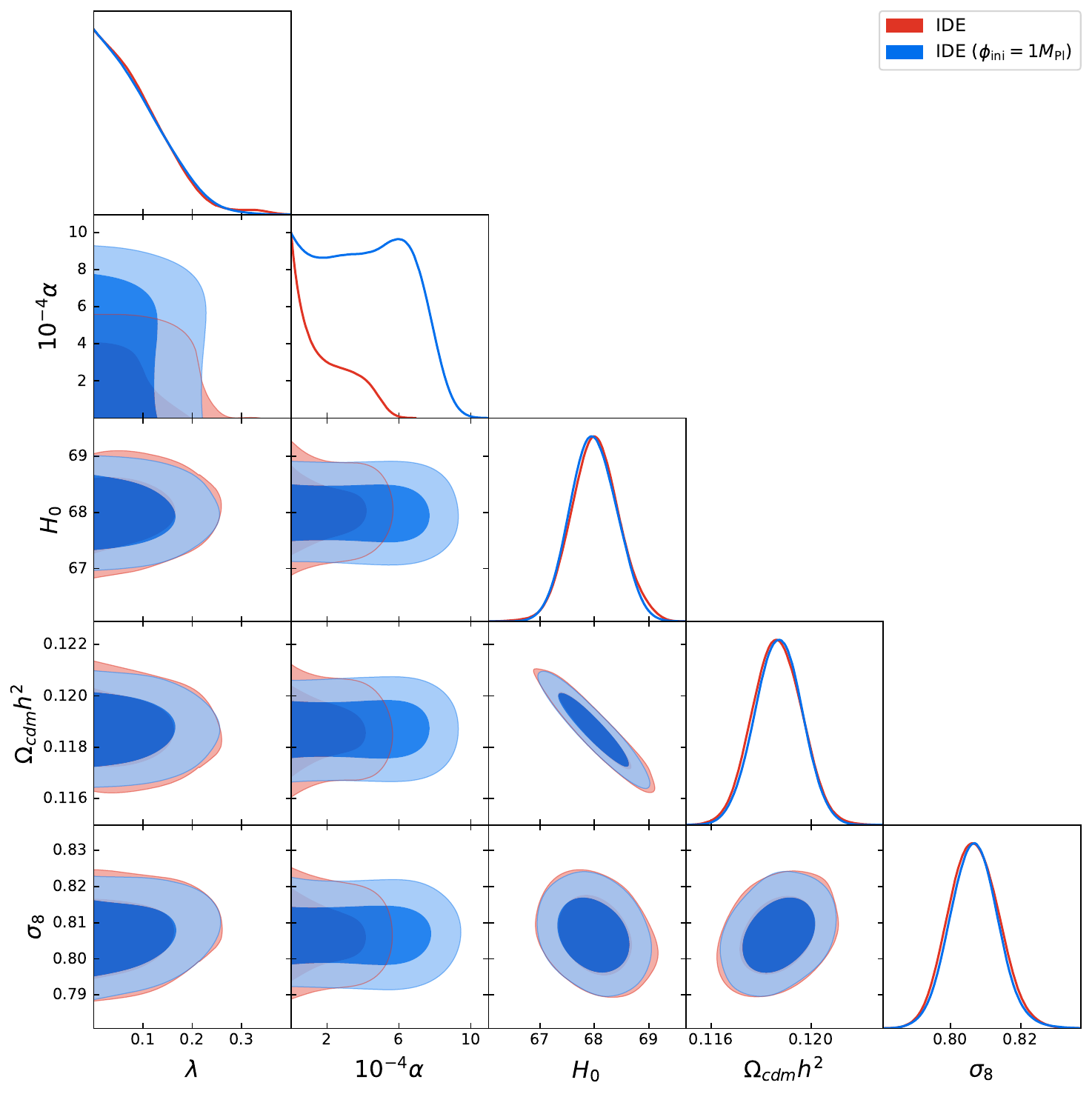}
    \caption{Marginalised posterior distributions for the IDE and IDE with fixed initial value of scalar field ($\phi_{\rm ini} = 1 M_{\rm Pl}$) models using PL18+BAO+Pantheon+CC+RSD.}
    \label{fig:poster_distributions_2d}
\end{figure*}

\begin{table*}[t!]
\caption{Observational constraints at a $68 \%$ confidence level on the independent and derived cosmological parameters for the $\Lambda$CDM, IDE, IDE with fixed $\phi_{\rm ini}$, and uncoupled quintessence models using PL18+BAO+Pantheon+CC+RSD. The quantities in the second half of this table are the derived parameters of our analysis which are the redshift of reionization $z_{reio}$, the Helium fraction $Y_{He}$, the Hubble constant $H_0$, the absolute magnitude of SN1a as inferred from the data sets used $M$, and the present day mass fluctuation amplitude at $8h^{-1} {\rm Mpc}$ $\sigma_8$.}
\begin{tabular}{C{3.5cm} C{3.5cm} C{3.5cm} C{3.5cm} C{3.5cm}}
\hline\hline \\[-1.5ex]
Parameter & $\Lambda${\rm CDM} & IDE & \begin{tabular}{@{}c@{}} IDE  \\ ($\phi_{\rm ini} =  1 M_{ \rm Pl}$) \end{tabular} & \begin{tabular}{@{}c@{}} Uncoupled quintessence   \\ ($\alpha = 0$) \end{tabular}  \\
\hline \\[-1.5ex]
$\Omega_b h^2$ & $0.02247\pm 0.00013$ &  $0.02247\pm 0.00013$ &  $0.02246\pm 0.00013        $&  $0.02248\pm 0.00013$\\
$\Omega_{cdm} h^2$ &   $0.11872\pm 0.00095$ &  $0.11864\pm 0.00098 $ &  $0.11869\pm 0.00093        $& $0.11856\pm 0.00096 $\\
$100 \theta_s$ &  $1.04200\pm 0.00029   $ & $1.04202\pm 0.00028  $ &  $1.04200\pm 0.00029        $ & $1.04201\pm 0.00028  $\\
$\ln \left( 10^{10} A_s \right)$ &  $3.041\pm 0.016 $ & $3.041\pm 0.016  $ &  $3.041\pm 0.016            $ & $3.042\pm 0.016  $\\
$n_s$ &   $0.9686\pm 0.0037 $ & $0.9687\pm 0.0039 $  &  $0.9685\pm 0.0036          $ & $0.9690\pm 0.0037 $\\
$\tau_{reio}$ &  $0.0543\pm 0.0078 $ & $0.0540\pm 0.0080 $  &  $0.0541\pm 0.0077          $ & $0.0545\pm 0.0076   $\\
$10^{-4} \alpha$ &  - &  $< 2.67   $  &  $< 5.73$ & - \\
$\lambda$ &  - & $< 0.109 $ &  $< 0.109                   $ & $< 0.406   $\\
$\phi_{\rm ini}$ & -  &  $1.0004^{+0.0078}_{-0.0089} $ &  - & - \\
\hline \\[-2ex]
$z_{reio}$ &  $7.63\pm 0.79$  &  $7.60\pm 0.82  $  &  $7.61\pm 0.77              $ & $7.64\pm 0.77$\\
$Y_{He}$ &   $0.247881\pm 0.000056  $ &  $0.247882\pm 0.000056  $ &  $0.247880\pm 0.000056      $& $0.247885\pm 0.000057 $\\
$H_0$ &  $67.98\pm 0.42 $ &   $68.02\pm 0.44 $ &  $67.99\pm 0.42             $ & $67.67^{+0.56}_{-0.45} $\\
$M$ & $-19.410\pm 0.012 $  & $-19.409\pm 0.012  $  & $-19.410\pm 0.012          $ & $-19.415\pm 0.013  $\\
$\sigma_8$ &  $0.8057\pm 0.0069   $ &   $0.8066\pm 0.0071 $ &  $0.8067\pm 0.0068          $ & $0.8019^{+0.0080}_{-0.0072}  $\\
[1ex]
\hline \\[-2ex]
%$\chi^{2}_{\rm min, i} - \chi^{2}_{\rm  min, \Lambda CDM}$ & - &   $xyz$  & $xyz$  \\[1ex]
$\ln B_{  i, \Lambda {\rm CDM}}$ & - & $-9.29$  & $-4.13$ & $-1.39$  \\[1ex]
\hline \\[-2ex]
\hline
\end{tabular}
\label{table:IDE_constraints}
\end{table*}

\begin{figure*}[t!]
    \centering
    \includegraphics[width=0.80\textwidth]{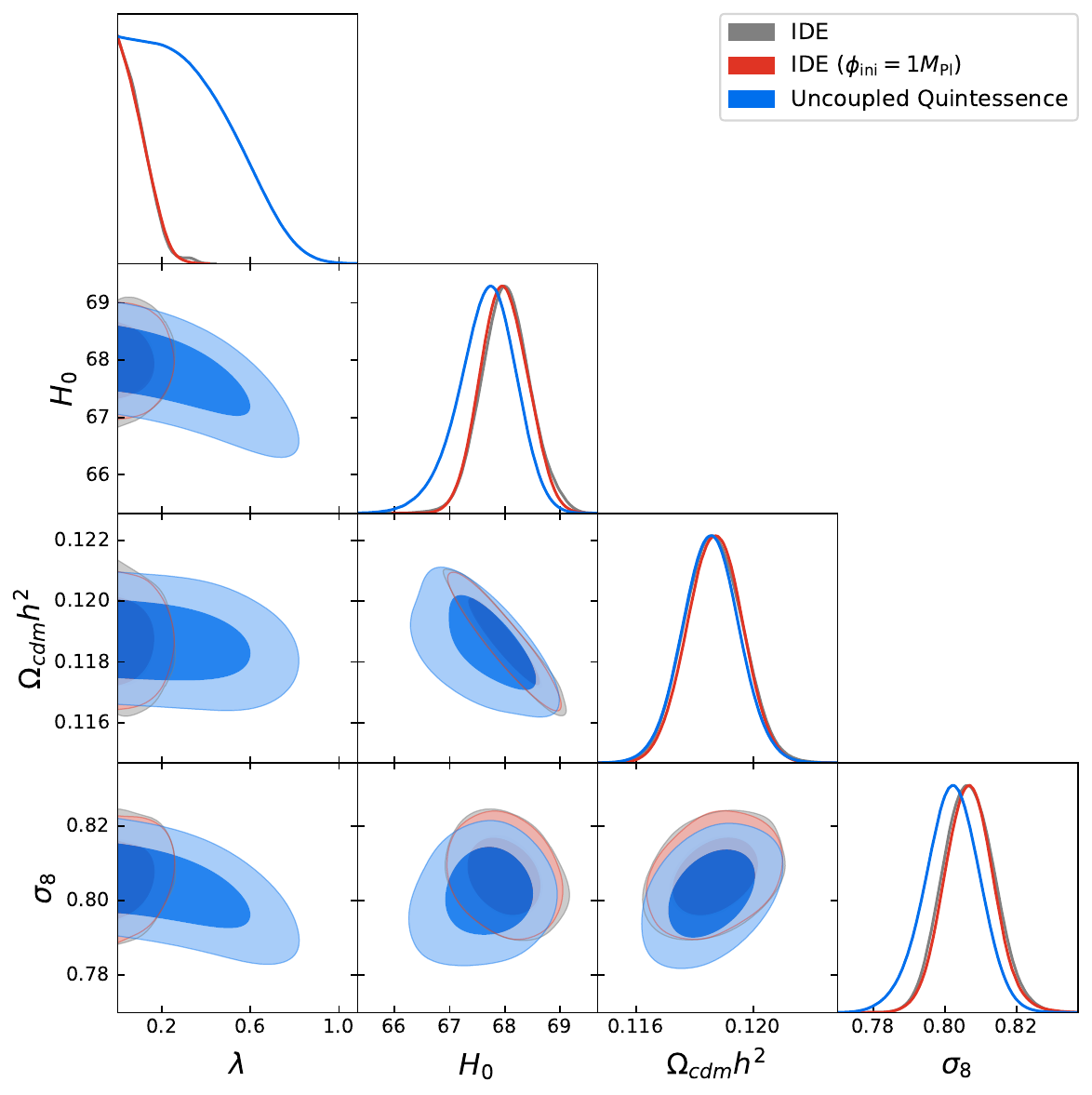}
    \caption{Marginalised posterior distributions for the IDE, IDE with fixed initial value of the scalar field ($\phi_{\rm ini} = 1 M_{\rm Pl}$), and the uncoupled quintessence with exponential potential ($\alpha=0$) models using PL18+BAO+Pantheon+CC+RSD.}
    \label{fig:poster_distributions_2d_alpha0}
\end{figure*}

In our analysis, the IDE model is described by a set of nine parameters whose priors are specified in \Cref{table:IDE_priors} and where $h$ is the reduced Hubble constant defined by $H_0 = 100 h {\rm km s^{-1} Mpc^{-1}}$. These parameters are the reduced baryon energy density $\Omega_b h^2$, the reduced CDM energy density $\Omega_{cdm} h^2$, the ratio of the sound horizon to the angular diameter distance at decoupling $\theta_s$, the scalar amplitude of the primordial power spectrum $A_s$, the scalar spectral index $n_s$, the reionization optical depth $\tau_{reio}$, the slope of the scalar field potential $\lambda$, the conformal coupling parameter $\alpha$, and the initial value of the scalar field $\phi_{\rm ini}$.

\begin{figure}[t!]
    \centering
    \includegraphics[width=0.30\textwidth]{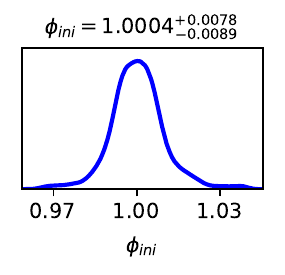}
    \caption{Marginalised posterior distribution for $\phi_{\rm ini}$ in the IDE model using PL18+BAO+Pantheon+CC+RSD.}
    \label{fig:phi_ini_distribution_1d}
\end{figure}

In order to numerically study the evolution of the background and cosmological perturbations for the Interacting Dark Energy (IDE) model described above we use a modified version of the $\texttt{CLASS}$\footnote{\href{https://github.com/lesgourg/class_public}{https://github.com/lesgourg/class\_public}} code \cite{lesgourgues2011cosmic, Blas_2011} to calculate the background evolution and the evolution of perturbations. For cosmological parameter exploration we use the Markov Chain Monte Carlo sampling package $\texttt{MontePython}$\footnote{\href{https://github.com/brinckmann/montepython_public}{https://github.com/brinckmann/montepython\_public}} \cite{Brinckmann:2018cvx, Audren:2012wb} in conjunction with the data sets outlined below. In addition to this, the $\texttt{GetDist}$\footnote{\href{https://github.com/cmbant/getdist}{https://github.com/cmbant/getdist}} \cite{Lewis:2019xzd} Python package is used to analyse the chains and produce the values and plots of the parameters in \Cref{table:IDE_constraints} and in \Cref{fig:poster_distributions_2d,fig:poster_distributions_2d_alpha0,fig:phi_ini_distribution_1d}. Finally, we calculate the Bayes factor of a scalar field model with respect to the $\Lambda$CDM model, $B_{i, \Lambda {\rm CDM}}$, by using the $\texttt{MCEvidence}$\footnote{\href{https://github.com/yabebalFantaye/MCEvidence}{https://github.com/yabebalFantaye/MCEvidence}} code \cite{Heavens:2017afc}. The natural logarithm of this Bayes factor, $\ln B_{  i, \Lambda {\rm CDM}}$, is shown in the last row of \Cref{table:IDE_constraints}.

We use the following combination of recent observational data sets in order to analyse and constrain the IDE model:

\begin{itemize}

\item{\textbf{Cosmic Microwave Background: }}

We use the full TTTEEE+lowE Cosmic Microwave Background (CMB) likelihood from the latest Planck 2018 release \cite{Planck:2018vyg}. This includes temperature (TT) and polarisation (EE) anisotropy data as well as cross-correlation data between temperature and polarisation (TE) at high and low multipoles.

\item{\textbf{Baryon Acoustic Oscillations: }}

We consider Baryon Acoustic Oscillations (BAO) measurements coming from  BOSS DR12 \cite{10.1093/mnras/stx721}, 6dFGS \cite{10.1111/j.1365-2966.2011.19250.x}, and SDSS-MGS  \cite{10.1093/mnras/stv154} for use in our analysis.

\item{\textbf{Type Ia Supernovae: }}

We use the Pantheon data catalog consisting of $1048$ points in the region $z \in [0.01, 2.3]$ of SNIa luminosity distance data as provided by \cite{Pan-STARRS1:2017jku}.

\item{\textbf{Redshift Space Distortions: }}

We employ the `Gold 2018' Redshift Space Distortions (RSD) data set compilation consisting of 22 measurements as described in \cite{Sagredo:2018ahx} and a likelihood code as detailed in \cite{1798956}.

\item{\textbf{Cosmic Chronometers: }}

We use 31 measurements of $H(z)$ from cosmic chronometers (CC) in the redshift range $z \in [0.07, 1.965]$ as detailed in Table 4 of \cite{Moresco_2016}.

\end{itemize}

The results for our data analysis are shown in \Cref{table:IDE_constraints} where we compare constraints of the $\Lambda$CDM, IDE, IDE with fixed $\phi_{\rm ini}$, and uncoupled quintessence models using the full PL18+BAO+Pantheon+CC+RSD data set for all models; in  \Cref{fig:poster_distributions_2d} where we compare the marginalised posterior distributions for parameters in the IDE and IDE with fixed $\phi_{\rm ini}$ models; in \Cref{fig:poster_distributions_2d_alpha0} where we compare the marginalised posterior distributions for parameters in the IDE, IDE with fixed $\phi_{\rm ini}$, and uncoupled quintessence models; and in \Cref{fig:phi_ini_distribution_1d} where we present the marginalised posterior distribution for $\phi_{\rm ini}$ in the IDE model. We utilise the full PL18+BAO+Pantheon+CC+RSD data set in all cases in order to obtain convergence for all the models, with all parameters achieving $|R-1|<0.03$, where $R$ is the Gelman-Rubin statistic \cite{10.1214/ss/1177011136}. Since attractor mechanisms drive the field towards the minimum of the coupling function in the early universe, we also investigate the case where the initial value of the scalar field sits directly at the minimum, i.e. when $\phi_{\rm ini} = 1 M_{\rm Pl}$, in this paper. In addition to $\Lambda$CDM and the IDE model, we also consider the uncoupled model (by imposing $\alpha = 0$) for comparison and to illustrate the impact of including the coupling on the parameter constraints. For the uncoupled model, owing to the form of the potential, we fix the initial value of the scalar field $\phi_{\rm ini}$ and do not allow it to vary.

One of the first things to notice is the large values of the coupling constant, $\alpha$, allowed by the data for the IDE models as can be seen in \Cref{fig:poster_distributions_2d}. We obtain an upper limit for $\alpha$ in both cases, of $10^{-4} \alpha < 2.67$ at $1 \sigma$ for the IDE model and $10^{-4} \alpha < 5.73$ at $1 \sigma$ for the IDE ($\phi_{\rm ini} =  1 M_{ \rm Pl}$) model. Larger values of $\alpha$ are allowed in the IDE ($\phi_{\rm ini} =  1 M_{ \rm Pl}$) model since $G_{\rm eff} = G_N$ until the late universe, whereas in the case where $\phi_{\rm ini}$ varies, $\alpha$ can be constrained through the fact that $G_{\rm eff} \neq G_N$ at the time simulations begin.

As can be seen in the upper panel of \Cref{fig:Geff_plot_alpha}, the large values of $\alpha$ supported by the data do not actually have as grand an impact on $G_{\rm eff}$ as one may initially believe. We can see that a ten-fold increase in $\alpha$ (from $10^{-4}\alpha = 0.01$ to $10^{-4}\alpha = 0.1$) leads to a $\sim 5 \%$ increase in $G_{\rm eff}$, but a ten-fold increase from $10^{-4}\alpha = 0.5$ to $10^{-4}\alpha = 5$ leads to only a sub-percent increase in  $G_{\rm eff}$, leaving $G_{\rm eff}$ with an overall enhancement of $\sim 14\%$ from $G_N$. $G_{\rm eff}$ is, however, quite sensitive to changes in the slope of the scalar field potential, $\lambda$, as can be seen in the lower panel of \Cref{fig:Geff_plot_alpha}. The reason for this is that $G_{\rm eff}$ depends on $\alpha$ {\it and} on the value of $\phi$. Larger values of $\alpha$ imply a steeper coupling function and as a consequence, the field is kept at the minimum more effectively. For all cases in the IDE models, we find that the field excursion is smaller than the Planck mass, i.e. ($|\phi - \phi_*| < M_{\rm Pl}$).

We obtain an upper limit on the slope of the potential of $\lambda < 0.109$ at $1 \sigma$ for the coupled IDE models and an upper limit of $\lambda < 0.406$ at $1 \sigma$ for the uncoupled quintessence model. We can see that this IDE model does not help to alleviate the swampland requirement of $|V'/V| \geq  c \sim \mathcal{O}(1)$ and in fact exacerbates the tension when compared to the uncoupled model. We also note that the coupled model breaks the degeneracy in the $(H_0,\lambda)$-plane and the $(\sigma_8,\lambda)$-plane, as it can be seen in  \Cref{fig:poster_distributions_2d_alpha0}.

In \Cref{fig:poster_distributions_2d}, we see that for the IDE model, there exists a negative correlation between the conformal coupling constant, $\alpha$, and slope of the scalar field potential, $\lambda$, with a higher value of $\alpha$ being compensated by a lower value of $\lambda$. This degeneracy does not exist when $\phi_{\rm ini}$ is set to the minimum of the coupling function.

We justify the prior for the initial value of the scalar field, $\phi_{\rm ini}$, to be centred around the minimum of the coupling function, $\phi_*$, owing to the attractor mechanisms aforementioned. We allow $\phi_{\rm ini}$ to vary as a cosmological parameter in our analysis for the IDE model and find that it is tightly constrained about the minimum of the coupling function with a mean of $\phi_{\rm ini}/M_{\rm Pl} = 1.0004^{+0.0078}_{-0.0089}$ at $1 \sigma$ as in \Cref{fig:phi_ini_distribution_1d}, in agreement with the attractor mechanisms discussed in \cite{van_de_Bruck_2019}.

We can see from \Cref{table:IDE_constraints} that the IDE and IDE ($\phi_{\rm ini} = 1 M_{\rm Pl}$) models are in strong agreement with each other and that the IDE models are also not in tension with the uncoupled quintessence model. In \Cref{table:IDE_constraints} we also see that the IDE model is in good agreement with $\Lambda$CDM for this data set combination, with no significant tensions observed in the cosmological parameters between the two models. This is to be expected as the model behaves similarly to uncoupled quintessence (with an exponential potential) up until $z \sim 1$, when the field switches on, at which point it begins to deviate from an uncoupled model.

In the last row of \Cref{table:IDE_constraints}, we see that $\ln B_{  i, \Lambda {\rm CDM}} < 0$ for both the coupled and uncoupled models. According to Jeffrey's scale \cite{doi:10.1080/01621459.1995.10476572}, this indicates that $\Lambda$CDM is preferred over the scalar field models. The introduction of an extra parameter in the uncoupled quintessence model, i.e. $\lambda$, is not enough to improve the fit over  $\Lambda$CDM, likewise, the introduction of two extra parameters in the IDE ($\phi_{\rm ini} = 1 M_{\rm Pl}$) model or three extra parameters in the IDE model do not improve the fit and in fact decrease the Bayes factor even further.

\section{Summary and Conclusions}
\label{Conclusions}

In this paper we have studied a specific IDE model, in which a fifth force between DM particles switches on at the onset of DE domination. This is achieved by considering a coupling function which possesses a minimum at a certain field value $\phi_*$. The DE field evolves away from the minimum due to the influence of the potential energy. In contrast to other interacting DE models, the fifth force between DM particles becomes important only at the end of matter domination, if the field does sit at the minimum of the coupling function initially. We used several data sets to constrain $\phi_{\rm ini}$, the coupling parameter $\alpha$ and the slope of the potential energy $\lambda$.

Our findings can be summarized as follows:
\begin{itemize}
    \item The initial value of the field, deep inside the radiation dominated epoch, is constrained to be very near the minimum of the coupling function. 
    \item Our best fit value for the slope $\lambda$ of the potential is lower than other IDE models, see e.g. \cite{VanDeBruck:2017mua}, although different data sets have been used in this study.  
    \item The coupling parameter $\alpha$ is only weakly constrained ($10^{-4}\alpha \lesssim 2.7$ at $1 \sigma$ for the  IDE model and $10^{-4}\alpha \lesssim 5.7$ at $1 \sigma$ for the IDE ($\phi_{\rm ini} =  1 M_{ \rm Pl}$) model).
    \item Since the effective coupling depends also on the value of the DE field $\phi$, the effective gravitational constant is determined by the dynamics of DE and hence by both $\alpha$ and $\lambda$. The dependence of the effective gravitational coupling on $\lambda$ and $\alpha$ is illustrated in \Cref{fig:Geff_plot_alpha}. 
\end{itemize}

Our study shows that equivalence principle violations in the dark sector at the present epoch are much less constrained than previously thought. This is because such new interactions may result from the dynamics of DE itself and only become important rather late in the cosmic history. The data sets used in this paper, which measure the evolution of the cosmological background and cosmological perturbations at the linear level, do not constrain the fifth force strongly and the model is, for a wide range of parameters, indistinguishable from the $\Lambda$CDM model. We do expect the model to differ from $\Lambda$CDM at much smaller length scales, for which linear perturbation theory is no longer adequate. The fifth force is likely to leave an imprint on the evolution of non-linear perturbations and affect, for example, the matter power spectra on small scales. It would also be interesting to study the tidal tail test \cite{Kesden:2006vz} in this model, since the coupling switches on only at late times. We do expect this test to constrain this model further. We leave these studies for future work. 

\section*{Acknowledgements}
We are grateful to Ed Copeland and Adam Moss for useful remarks during the early stages of this work. CCT is supported by a studentship from the School of Mathematics and Statistics at the University of Sheffield. CvdB is supported (in part) by the Lancaster–Manchester–Sheffield Consortium for Fundamental Physics under STFC grant: ST/T001038/1. 

\clearpage

\bibliography{references}

\end{document}